\documentstyle[floats,twocolumn,prl,aps,psfig]{revtex} 
\newcommand{\be}{\begin{equation}}
\newcommand{\ee}{\end{equation}}
\newcommand{\bea}{\begin{eqnarray}}
\newcommand{\eea}{\end{eqnarray}}

\begin{document}
\draft
\title{Phase Diagram of the Extended Hubbard Model with Pair Hopping 
Interaction}

\author{G.I. Japaridze $^{a,b}$ and Sujit Sarkar $^{a}$}
\address{
$^a$ Max-Planck-Institut f$\ddot{u}$r Physik of komplexer Systeme,\\ 
N$\ddot{o}$thnitzer Str.\ 38, D-01187, Dresden, Germany.\\
$^b$ Institute of Physics, Georgian Academy of Sciences,
Tamarashvili Str.\ 6, 380077, Tbilisi, Georgia.}

\address{~
\parbox{14cm}{\rm 
\medskip
A one-dimensional model of interacting electrons with on-site $U$, 
nearest-neighbor $V$, and pair-hopping interaction $W$ is 
studied at half-filling using the continuum limit field theory approach. The 
ground state phase diagram is obtained for a wide range of coupling constants. 
In addition to the insulating spin- and charge-density wave phases for large 
$U$ and $V$, respectively, we identify bond-located ordered phases 
corresponding to an enhanced Peierls instability in the system for 
$W<0$, $|U-2V|<|2W|$, and to a staggered magnetization located 
on bonds between sites for $W>0$, $|U-2V|<W$. The 
general ground state phase diagram including insulating, metallic, and 
superconducting phases is discussed. A transition to the $\eta_{\pi}$-superconducting 
phase at $|U-2V| \ll 2t \leq W$ is briefly discussed.
}}
\maketitle

\section{\bf Introduction}

The one dimensional (1D) extended Hubbard model with nearest-neighbor repulsion $V$, in addition 
to the on-site repulsion $U$ (hereafter $U-V$ model) has been extensively studied during the 
last two decades as an important theoretical test-bed for studying low-dimensional strongly 
correlated electron systems with rich phase structures.  Considerable attention has been
focused on stuying 
of the ground state (GS) phase diagram of the $U-V$ model at half-filling, using analytical studies 
and numerical simulations
\cite{Emery1,Solyom,Hirsch1,Hirsch2,CanFrad,CSF,Voit,Dongen,Zg,Nakamura1,DKC1,Nakamura2,DKC2,Nakamura3,Furusaki}
The sketch of the phase diagram 
consists of a Mott insulating phase ($U>2|V|$) with dominating spin-density wave correlations, 
an insulating long-range-ordered (LRO) charge-density-wave (CDW) 
phase ($2V>U>0$), and metallic 
phases with dominating singlet (SS) and triplet 
(TS) superconducting correlations. In the 
physically most interesting region of repulsive interactions ($U,V>0$), the weak-coupling 
perturbative renormalization group 
studies  \cite{Emery1,Solyom} show that there is a continuous 
phase transition between SDW and CDW along the line $U=2V$. In the strong coupling limit 
($U,V >> 1$) the SDW-CDW transition is discontinuous (first order) and the phase boundary is slightly 
shifted away from the line $U=2V$  \cite{Hirsch1,Hirsch2,Dongen}.
Estimates for the location of the 
tricritical point, where the nature of the transition changes, have ranged from 
$U_{c} \simeq 1.5$ to $U_{c} \simeq 5$ (and $V_{c} \simeq U_{c}/2$). 
\cite{Hirsch1,Hirsch2,CanFrad,Zg,DKC2} Recently increased interest towards the $U-V$ Hubbard 
model was triggered by Nakamura,\cite{Nakamura1,DKC1,Nakamura2} who found numerically that 
for small to intermediate values of $U$ and $V$, the SDW and CDW phases are mediated by the 
bond-ordered charge-density-wave (BO-CDW) phase. The SDW-CDW transition splits into 
two separate transitions: ({\it i}) a Kosterlitz-Thouless spin gap transition from SDW to  BO-CDW 
and  ({\it ii}) a continuous transition from BO-CDW to CDW.

An analogous sequence of phase transitions in the vicinity of the $U=2V$ line
is the intrinsic feature of  
extended $U-V$ Hubbard models with bond-charge coupling \cite{Jap1,JK}. The {\it bond 
located ordering} in these models is directly connected with the {\em site-off-diagonal nature} of 
the bond-charge coupling. Models of correlated electrons with bond-charge
coupling have currently attracted 
a great interest as models showing unconventional, ``kinematical''  mechanisms of 
superconducting correlations
\cite{Hirsch3,EKS,ES2,JM2,ES3,ES4,ES5,Refes5,ES6,PK,AM,hui,Refes4,SA,BR,BC,Refes6,BJ,JM1,Rob1,Rob2,JKSKB}.
Among the models with correlated ``kinematics'' 
models with pair-hopping interaction are the subject of current studies 
\cite{PK,AM,hui,Refes4,SA,BR,BC,Refes6,BJ,JM1,Rob1,Rob2,JKSKB}. 
In this paper we consider the ground state phase diagram of extended $U-V$ Hubbard model 
supplemented with the pair hopping term. The Hamiltonian of the model is given by
\begin{eqnarray} 
{\cal H}  & = & -t \sum_{n,\sigma}(c^{\dagger}_{n, \sigma}c_{n+1,\sigma} 
+ c^{\dagger}_{n+1,\sigma}c_{n, \sigma}) 
- \mu \sum_{n,\sigma}c^{\dagger}_{n, \sigma}c_{n, \sigma}\nonumber\\ 
& + & {1 \over 2}U \sum_{n,\sigma}\hat \rho_{n,\sigma}\hat \rho_{n,-\sigma} 
+ V\sum_{n}\hat \rho_{n}\hat \rho_{n+1}\nonumber\\
& + & W\sum_{n}(c^{\dagger}_{n, \uparrow} c^{\dagger}_{n, \downarrow}
c_{n+1, \downarrow} c_{n+1, \uparrow}+ h.c)\label{UVWhamiltonian},
\end{eqnarray}
where $\hat \rho_{n,\sigma}= c^{\dagger}_{n, \sigma}c_{n,\sigma}$, 
$\hat \rho_{n} = \sum_{\sigma} \hat \rho_{n,\sigma}$, and $c^{\dagger}_{n, \sigma}$ 
 ($c_{n,\sigma }$) denotes  the creation (annihilation) operator for an 
electron with spin $\sigma $ at site $n$. In Eq.\ (\ref{UVWhamiltonian}), 
$t$ and $\mu$ denote the hopping integral and the chemical potential respectively,
with $U$ being the on-site Coulomb-Hubbard repulsion and $V$ the intersite interaction.
$W$ is the pair hopping interaction. 

It is notable that the $U$, $V$, and $W$ terms could be 
obtained from the same general tight-binding Hamiltonian \cite{HUB} by focusing on a selected 
term of the two-particle interaction. The sign of the {\em Coulomb-driven} coupling constants is 
typically repulsive $U,V,W>0$, and usually $W \ll U,V$.  However, below we will treat these 
parameters as the effective (phenomenological) ones, assuming that they include all the 
possible renormalizations, and their values and signs could be arbitrary.

Interest in models with pair-hopping coupling comes from the unusual mechanisms of 
Cooper pairing provided by this interaction. In the absence of the on-site and nearest-neighbor 
couplings ($U=V=0$), the model Eq.\ (\ref{UVWhamiltonian}) reduces to the Penson-Kolb (PK) model 
\cite{PK}.  The PK model is  possible the simplest model which captures the essential physics of 
an electron system showing the $\eta$-superconductivity in the ground state. In
the $\eta$-paired state, 
the eigenstates of the correlated electrons are constructed 
exclusively in terms of doublon (on-site singlet 
pair) creation operators \cite{CNY1}. We consider two different realizations of the $\eta$-paired state, 
constructed in terms of zero size Cooper pairs with {\it center-of-mass momentum equal to} zero 
($\eta_{0}$-pairing) and $\pi$ ($\eta_{\pi}$-pairing), respectively. 

In the case of an ``attractive'' ($W<0$) pair-hopping interaction the PK model describes a 
{\em continuous evolution} of the usual BCS type superconducting state at 
$\left|U\right|,\left|W\right| \ll t$ into a local pair $\eta_{0}$-superconducting 
state at $\left|W\right|/t \rightarrow \infty $ \cite{AM}.  In the case of repulsive ($W>0$) pair-hopping 
interaction, in contrast, the transition into the $\eta_{\pi}$-paired state takes place at 
{\em finite} $W_{c}$ and is of {\em first order} (level-crossing type) \cite{SA,BJ,JKSKB}. 

In this paper we address the question, whether the pair-hopping coupling could 
lead to the superconducting ordering in the physically most relevant region of 
parameters $U,V \gg W>0$. In this communication we present the weak-coupling ground state 
phase diagram of the model Eq.\ (\ref{UVWhamiltonian}). As we show in this paper, the  ``attractive'' 
($W > 0$) pair-hopping coupling enlarges the region of 
coupling constant corresponding to the metallic 
phase with dominating SS and TS instabilities. However, in the repulsive sector of the phase diagram, 
 along the line $U=2V>0$, only the insulating LRO BO-CDW phase, is realized at $|U-2V|<|W|$. 
In the case of a ``repulsive" ($W>0$) pair-hopping coupling, the BO-SDW phase corresponding to a 
bond located staggered magnetization is together with the $CDW$ the most divergent instability 
in the system. We also present quantitative arguments in favour of an additional phase transition 
at $W \simeq 4t$ from the insulating BO-SDW to the $\eta_{\pi}$-superconducting phase. 

The outline of the paper is as follows: In section II we present the
continuum limit bosonized version of the model. In section III we
discuss the weak coupling phase diagram. Section IV is devoted to
a discussion of the ground state phase diagram and a summary.

\section{\bf Continuum limit theory and bosonization.}

In this section we derive the low-energy effective field theory of the lattice 
model Eq.\ (1) at half-filling. Considering the weak-coupling case $|U|, |V|,|W|  \ll t$ 
, we linearize the spectrum and pass to the continuum limit by use of the mapping 
\begin{equation}
a_{0}^{-1/2}c_{n,\sigma }\rightarrow {\it i}^{n}R_{\sigma}(x) + 
(-{\it i})^{n}L_{\sigma }(x).  \label{linearization}
\end{equation}
Here $x=na_{0}$, $a_{0}$ is the lattice spacing, and $R_{\sigma}(x)$ and 
$L_{\sigma}(x)$ describe right-moving and left-moving particles, respectively. 
These fields can be bosonized in a standard way \cite{go}:
\bea
R_{\sigma}(x)&=&(2\pi a_{0})^{-1/2}e^{{\it i}\sqrt{4\pi}\Phi_{R,\sigma}(x)},\\
L_{\sigma}(x)&=&(2\pi a_{0})^{-1/2}e^{-{\it i}\sqrt{4\pi}\Phi_{L,\sigma}(x)},
\eea
where $\Phi_{R(L),\sigma}(x)$ are the right (left) moving Bose fields.  We define 
$\Phi_{\sigma}=\Phi_{R,\sigma}+\Phi_{L,\sigma}$ and introduce linear combinations, 
$\varphi_{c}=(\Phi_{\uparrow}+ \Phi_{\downarrow})/\sqrt{2}$ and 
$\varphi_{s}=(\Phi_{\uparrow}-\Phi_{\downarrow})/\sqrt{2}$, to 
describe the charge and spin degrees of freedom, respectively. Then, after a 
rescaling of fields and 
lengths,  we rewrite the  bosonized version of the Hamiltonian (\ref{UVWhamiltonian}) 
in terms of two decoupled quantum SG theories, ${\cal H}= {\cal H}_{c}+{\cal H}_{s}$, where
\begin{eqnarray}
{\cal H}_{c(s)}&=&\int dx\Big\{{\frac{v_{c(s)}}{2}}\left[(\partial _{x}\varphi
_{c(s)})^{2}+(\partial _{x}\vartheta_{c,(s)})^{2}\right]\nonumber\\
&+&{\frac{m_{c(s)}}{\pi a _{0}^{2}
}}\cos \big(\sqrt{8\pi K_{c(s)}}\varphi _{c}(x)\big)\Big\}.  \label{SG}
\end{eqnarray}
Here $\theta _{c(s)}(x)$ are the dual counterparts of the fields $\phi _{c(s)}(x)$: 
$\partial _{x}\theta _{c(s)}=\Pi _{c(s)}$ where $\Pi _{c(s)}$ is the momentum 
conjugate to the field $\phi _{c(s)}$.  Here we have defined
\begin{eqnarray}
K_{c} &=&(1-g_{c})^{-1/2}\simeq 1+{\frac{1}{2}}g_{c},\hskip0.2cm m_{c}=-\frac{
g_{u}}{2\pi },  \label{Kc} \\
K_{s} &=&(1-g_{s})^{-1/2}\simeq 1+{\frac{1}{2}}g_{s},\hskip0.2cm m_{s}=\frac{
g_{\perp }}{2\pi },  \label{Ks} 
\end{eqnarray}
$v_{c(s)}=v_{F}K^{-1}_{c(s)}$ are the velocities of the charge and spin excitations, 
$v_{F}= 2ta_{0}(1-W/\pi t)$, and  the small dimensionless coupling constants are given by 
\bea\label{Param1}
g_{s}=g_{\perp }&=&(U-2V+2W)/\pi v_{F},\\  
g_{c}&=&-(U+6V+2W)/\pi v_{F},\label{Param2}\\  
g_{u}&=&(U-2V-2W)/\pi v_{F}.  \label{Param3}
\eea
The relation between $K_{c}$ ($K_{s}$), $m_{c}$ ($m_{s}$), and $g_{c}$ 
($g_{s}$), $g_{u}$ ($g_{\perp}$) is universal in the weak coupling limit. 

In obtaining (\ref{SG}) the strongly irrelevant term
$\sim$ $\cos(\sqrt{8\pi K_{c}}\varphi_{c})\cos(\sqrt{8\pi K_{s}}\varphi_{s})$ 
describing umklapp scattering processes with parallel spins was omitted. 
The mapping of the Hamiltonian (\ref{UVWhamiltonian}) onto the quantum theory of two 
independent charge and spin Bose fields, allows a study of the ground state phase diagram of the 
initial electron  system using the far-infrared properties of the bosonic Hamiltonians (\ref{SG}). 
Depending on the relation between the bare coupling constants $K$ and $m$ the infrared 
behavior of the quantum SG field exhibits two different regimes \cite{Wieg}:

For $\left|m \right| \leq 2(K-1)$ we are in the weak
coupling regime; the effective mass $M \rightarrow 0$. The low energy
(large distance) behavior of the gapless charge (spin) excitations is
described by a free scalar field. The corresponding correlations show a
power law decay 
\begin{eqnarray}
\langle e^{i\sqrt{2\pi K^{\ast }}\varphi (x)}e^{-i\sqrt{2\pi K^{\ast }}
\varphi (x^{\prime })}\rangle &\sim &\left| x-x^{\prime }\right| ^{-K^{\ast
}},  \label{freecorrelations1} \\
\langle e^{i\sqrt{2\pi /K^{\ast }}\theta (x)}e^{-i\sqrt{2\pi /K^{\ast }}
\theta (x^{\prime })}\rangle &\sim &\left| x-x^{\prime }\right| ^{-1/K^{\ast
}},  \label{freecorrelations2}
\end{eqnarray}
and the only parameter controlling the infrared behavior in the gapless
regime is the fixed-point value of the effective coupling constants $K_{c(s)}^{\ast}$.

For $\left|m \right| > 2(K-1)$ the
system scales to a strong coupling regime: Depending on the sign of the
bare mass $m$, the effective mass $M \rightarrow \pm \infty $,
which signals the crossover into a strong coupling regime and indicates the
dynamical generation of a commensurability gap in the excitation spectrum. 
The field $\varphi_{c(s)}$ gets ordered with the vacuum expectation values \cite{mut}
\begin{equation}
\langle \varphi_{c(s)} \rangle =\left\{ 
\begin{array}{l}
\sqrt{\displaystyle{\pi/8K_{c(s)}}}\hskip0.3cm (m > 0) \\ 
0\hskip1.4cm(m < 0)
\end{array}
\right. \,.  \label{orderedfields}
\end{equation}
The ordering of these fields determines the symmetry properties of the
possible ordered ground states of the fermionic system.

Using Eqs.\ (\ref{Param1})-(\ref{Param3}) and (\ref{orderedfields}),
one easily finds that there is a {\em gap in the spin excitation spectrum} 
$(M_{s}\rightarrow -\infty $) for 
$$
U-2V+2W<0.
$$
In this sector of coupling constants,
the $\varphi _{s}$ field gets ordered with vacuum expectation value 
$\langle\varphi _{s}\rangle =0$. At $U-2V+2W \geq 0$ the spin 
excitations are gapless and the low-energy properties =
of the spin sector are described by the free Bose field system with the 
fixed-point value of the parameter $K^{\ast}_{s}=1$. 

The charge sector is gapped for 
$$
U > \max \{2V+2W, -2|V| \}
$$
and for 
$$
U <2V+2W \hskip0.3cm {\mbox {\rm but}} \hskip0.3cm 2V+W>0. 
$$
In the former case $M_{c} \rightarrow -\infty $ and the vacuum expectation value of the charge field 
$\langle\varphi _{s}\rangle =0$, while in the latter case $M_{c} \rightarrow \infty $ and 
$\langle\varphi _{s}\rangle =\sqrt{\displaystyle{\pi/8K_{c}}}$.

In the sectors of coupling constants corresponding to the gapless charge excitation spectrum  
the properties of the charge degrees of freedom are described by the free Bose field 
$$
{\cal H}_{c}=\frac{v_c}{2}\left[K^{\ast}_{c}(\partial _{x}\varphi_{c})^{2} + 
\frac{1}{K^{\ast}_{c}}(\partial _{x}\vartheta_{c})^{2}\right], 
$$
with the fixed-point value of the parameter 
\begin{equation}
K^{\ast}_{c} \simeq 1 + \sqrt{2(U+2V)(W +2V)}/\pi v_{F}.
\end{equation}
Especially important
is the line $U=2V+2W$  corresponding to the fixed-point line $m_{c}=0, K_{c}-1<0$. 
Here the infrared properties of the gapless charge sector are described by a free 
massless Bose field with the bare value of the Luttinger liquid parameter 
$K_{c}$. 

To clarify the symmetry properties of the ground states of the system in
different sectors we introduce the following set of order parameters
describing the short wavelength fluctuations of the 

${\bullet}$ {\it site}-located charge and spin density:
\begin{eqnarray}\label{CDWop}
\Delta_{\small CDW} & = & (-1)^{n} \sum_{\sigma}\rho_{n, \sigma}\nonumber\\
& \sim & \sin(\sqrt{2\pi K_{c}}\varphi_{c}) 
\cos( \sqrt{2\pi K_{s}}\varphi_{s}) \,\\
\Delta_{\small SDW} & = & {(-1)}^n \sum_{\sigma}\sigma \rho_{n, \sigma}\nonumber\\
& \sim & \cos(\sqrt{2\pi K_{c}}\varphi_{c})
\sin(\sqrt{2\pi K_{s}}\varphi_{s}) \, ,\label{SDWop}   
\end{eqnarray}

${\bullet}$ {\it bond}-located charge--density:  
\begin{eqnarray}
\Delta_{\small BO-CDW} &  = &  (-1)^{n}
 \sum_{\sigma}(c^{\dagger}_{n, \sigma}c_{n+1,\sigma} + h.c.)\nonumber\\ 
& \sim  & \cos(\sqrt{2\pi K_{c}}\varphi_{c})  
\cos(\sqrt{2\pi K_{s}}\varphi_{s})\label{b-CDW} 
\end{eqnarray}
${\bullet}$ The {\it bond}-located spin--density:
\begin{eqnarray}
\Delta_{\small BO-SDW} &  = &  (-1)^{n} \sum_{\sigma}
\sigma (c^{\dagger}_{n, \sigma}c_{n+1,\sigma} + h.c.)\nonumber\\ 
& \sim  & \sin(\sqrt{2\pi K_{c}}\varphi_{c})
\sin(\sqrt{2\pi K_{s}}\varphi_{s}).\label{b-SDW} 
\end{eqnarray}

In addition we consider two superconducting order parameters corresponding to the 

${\bullet}$ singlet and triplet superconductivity:
\begin{eqnarray}
\Delta_{SS}(x) & = &R^{\dagger}_{\uparrow}(x)L^{\dagger}_{\downarrow}(x)  
- R^{\dagger}_{\downarrow}(x)L^{\dagger}_{\uparrow}(x) \nonumber\\
& \sim & \exp(i \sqrt{\frac{2\pi}{K_{c}}}\theta_{c}) 
\cos(\sqrt{2 \pi K_{s}}\varphi_{s}),\label{SSop}\\
\Delta_{TS}(x) & = & R^{\dagger}_{\uparrow}(x)L^{\dagger}_{\downarrow}(x)  + 
R^{\dagger}_{\downarrow}(x)L^{\dagger}_{\uparrow}(x) \nonumber\\
& \sim & \exp(i  \sqrt{\frac{2\pi}{K_{c}}}\theta_{c}) 
\sin(\sqrt{2 \pi K_{s}}\varphi_{s}).\label{TSop}
\end{eqnarray}

\section{Weak-coupling phase diagram}

With these results for the excitation spectrum and 
the behavior of the corresponding fields, 
Eqs.\ (\ref{freecorrelations1})-(\ref{orderedfields}), we now discuss the 
{\em weak-coupling} ground state phase diagram of the model (\ref{UVWhamiltonian}). 
Below we will focus on the new phases appearing in the phase-diagram due to the 
effect of the pair-hopping coupling. The phase diagram consists of 5 sectors  
(see Fig.\ 1 and Fig.\ 2). 
Sectors $A, B, C1,C2$ are present in the phase diagram of the  $U-V$  Hubbard model \cite{Voit}.

Sector A 
\begin{itemize}
\item   $U > \max \{2V+2W, -2|V| \}$,
\end{itemize}
corresponds to the ordinary Mott insulating phase:  The charge excitation spectrum is gapped, the 
spin sector is gapless. The ordering of the field $\varphi_{c}$ 
with vacuum expectation value 
$\langle \varphi_{c} \rangle = 0$ leads to a supression of the superconducting, 
CDW, and BO-SDW correlations. The SDW and Dimer correlations show a power-law 
decay at large distances
\begin{eqnarray}  \label{correlSDW+Dimer}
\langle \Delta_{{\small SDW}}(x) \Delta_{{\small SDW}}(x^{\prime})
\rangle &\sim & \langle\Delta_{{\small BO-CDW}}(x)\Delta_{{\small BO-CDW}
}(x^{\prime})\rangle  \nonumber \\
& \sim & \left| x - x^{\prime}\right|^{-1}.
\end{eqnarray}

Sector B
\begin{itemize}
\item $U < 2V-2|W|$ and $2V+W>0$
\end{itemize}
corresponds to the long-range ordered CDW insulating phase. The charge and spin excitations 
are gapped. The fields $\varphi_{c(s)}$ get ordered with vacuum expectation values  
$\langle \varphi_{s}\rangle  = 0$ and 
$\langle \varphi_{c} \rangle =\sqrt{\displaystyle{\pi/8K_{c}}}$ 
, and 
\begin{equation}
\langle \Delta_{\small CDW}(x)\Delta_{\small CDW}(x^{\prime })\rangle \sim {\rm 
constant}.
\end{equation}

Sector C1 
\begin{itemize}
\item  $U < \min\{2V -2W;-2V\} \quad {\mbox {\rm and}} \quad 2V + W <  0 $
\end{itemize}
corresponds to the Singlet Superconducting (SS) phase. There a gap exists in the spin excitation 
spectrum and the spin field is ordered with $\langle \varphi _{s}\rangle =0$.  The charge excitation 
spectrum is gapless with the fixed point walue of the parameter $K^{\ast}_{c}>1$. The SDW, 
BO-SDW, and the TS instabilities are suppressed. The CDW, BO-SDW, and the SS instabilities 
show a power-law decay at large distancesr. However since $K^{\ast}_{c}>1$ the  SS instability 
\begin{equation}
\langle \Delta_{\small SS}(x)\Delta_{\small SS}(x^{\prime })\rangle 
\sim \left| x-x^{\prime }\right|^{-1/ K_{c}}
\end{equation}
dominates in the ground state.

Sector C2
\begin{itemize}
\item  $-2|V|<U < 2V -2W \quad {\mbox {\rm and}} \quad  2V + W <  0$ 
\end{itemize}
corresponds to the Luttinger liquid phase with dominating superconducting instabilities. 
None of the conditions of charge and spin gap is satisfied
here. In this sector, the system shows the properties of a Luttinger liquid
with dominating superconducting instabilities TS and SS.
The singlet superconducting and triplet superconducting correlations      
show the same power law decay at large distances and the TS instability
dominates because of the weak logarithmic corrections\cite{Voit}. 

Finally we analyze the sectors describing the new phases. These new phases essentially 
appear along the SDW-CDW transition line $U=2V>0$ of the $U-V$ Hubbard model. 
In the weak-coupling limit, the transition from the Mott insulating phase at $U>2V$ to 
the CDW insulator at $U<2V$ is mediated by the Luttinger liquid phase with 
gapless spin and charge excitations.  AT $U=2V$ the Mott insulator charge gap 
closes and at $U-2V < 0$ the charge and the spin gap opens simultaneously. 
In the very presence 
of the pair-hopping interaction, the SDW-CDW transition splits into two 
transitions: Along the line 
$U=2V-2W$ the spin gap opens, while at $U=2V+2W$ the Mott insulator charge gap closes, and 
for $U< 2V+2W$ the CDW charge gap opens. In the case of an attractive pair-hopping interaction 
$W<0$ (Fig. 1) the spin gap opens in the presence of a Mott insulator charge gap.  Therefore, 
in sector D
\begin{itemize}
\item  $|U-2V|<2|W|  \quad {\mbox {\and}} \quad    2V+W>0$,
\end{itemize}
the charge and spin channels are gapped and both, charge and spin fields are ordered, 
$\langle\varphi_{c}\rangle = \langle \varphi_{s} \rangle =0$. In this case 
the long-range ordered $BOW$ phase 
\begin{equation}\label{correl2}
\langle \Delta_{\small BO-CDW}(x) \Delta_{\small BO-CDW}(x')\rangle \sim 
{\rm constant}
\end{equation}
is realized in the ground state.
\begin{figure}[t] 
\vspace{0mm}
\psfig{figure=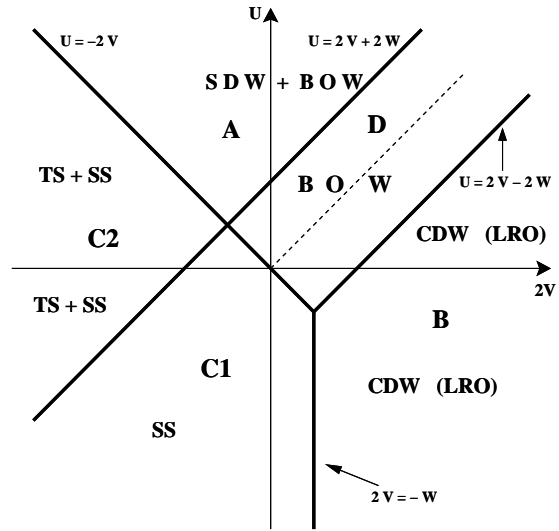 ,height=7.cm} 
\vspace{10mm}
\caption{The ground state phase diagram of the 1D $U-V-W$ model for the case of a 
half-filled band and  $W<0$. Solid lines separate different phases: 
A. $SDW + BOW$ - Mott insulating phase with an identical power-law 
decay of spin-density-wave and Peierls correlations. B.  CDW (LRO)-
long range ordered (LRO) charge density wave phase; C1. Singlet superconducting phase.
C2. Metallic phase with dominating singlet and triplet superconducting 
correlations. D. LRO dimerized (Peierls) phase. }
\label{fig1}
\end{figure} 

In the case of a repulsive pair-hopping coupling $W>0$ (Fig.\ 2), the transition within the charge 
degrees of freedom, takes place before the spin gap opens. Therefore, in sector D1
\begin{itemize}
\item  $|U-2V| <2|W| \quad {\mbox {\rm and}} \quad     U+2V>0$,
\end{itemize}
the generation of a gap in the charge excitation spectrum, 
accompanied by the ordering of the field $\varphi_{c}$ with vacuum 
expectation value $\langle \varphi_{c} \rangle = \sqrt{\displaystyle{\pi/8K_{c}}}$ 
, leads to a supression of the superconducting, SDW, and BO-CDW ordering. 
The CDW and BO-SDW correlations show a power-law decay at large distances
\begin{eqnarray}  \label{correlCDW+Bd-SDW}
\langle \Delta_{{\small CDW}}(x) \Delta_{{\small CDW}}(x^{\prime})
\rangle &\sim & \langle\Delta_{{\small BO-SDW}}(x)\Delta_{{\small BO-SDW}
}(x^{\prime})\rangle  \nonumber \\
& \sim & \left| x - x^{\prime}\right|^{-1}.
\end{eqnarray}
Therefore, this sector of the phase diagram corresponds to the insulating phase 
with coexisting CDW and BO-SDW instabilities.
\begin{figure}[t] 
\vspace{0mm}
\psfig{figure=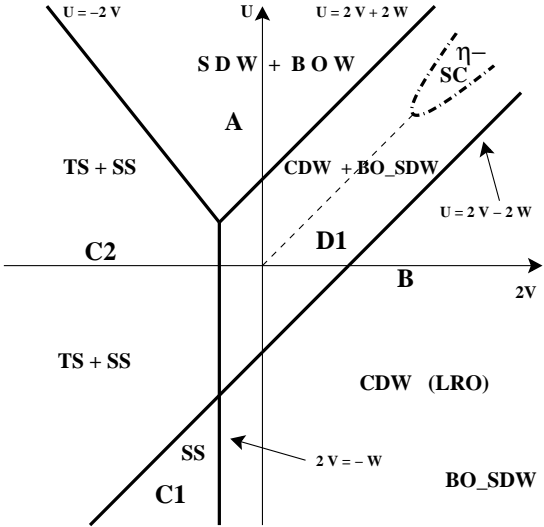 ,height=7.cm} 
\vspace{10mm}
\caption{The ground state phase diagram of the 1D $U-V-W$ model for the case of a 
half-filled band and  $W>0$. The dot-dashed line marks a transition to the 
$\eta_{\pi}$-superconducting phase.} 
\label{fig2}
\end{figure} 

Let us now discuss the $\eta_{\pi}$-superconducting phase. In models with ``kinematical'' 
mechanisms of Cooper pairing, transition to an $\eta$-paired phase is typically the 
{\em finite-bandwidth  phenomenon} \cite{EKS,ES3,ES4,ES5}. In the case of pair-hopping 
interaction,
the transition point is determined by the competition between the single-electron and 
doublon delocalization energies. After the transition the 
contribution of the one-electron hopping term 
to the ground state energy almost vanishes and the ground state energy is determined by the created 
strongly correlated two-particle $\eta_{\pi}$-pair band \cite{BJ,JKSKB}. Simultanously, after 
the transition the spin gap opens in the system while the charge gap (at half-filling) closes \cite{SA}.
In the case of the 
PK model the transition point  $W_{c}(U=V=0) \simeq 1.8t$  \cite{BJ}, while in the case of 
the on-site Hubbard repulsion, 
$W_{c}(V=0) \simeq 1.8t + \alpha U$, where  $\alpha $ is of the order of 
unity \cite{JKSKB}. In both cases the insulating CDW +(BO-SDW) phase is unstable toward 
transition to the $\eta_{\pi}$-superconducting state \cite{JM1,Rob1,JKSKB}. Due to the 
finite-bandwidth  nature of the transition to a $\eta_{\pi}$-paired state, 
it could not be consistently 
studied within the continuum-limit (infinite band) approach. 
Nevertheless, the existence of a transition is 
clearly traced in the additive renormalization of the Fermi velocity (bandwidth) by the pair-hopping term 
$v_{F}= 2ta_{0}(1-W/\pi t)$.  
In the narrow stripe along the frustration line $|U-2V| \ll W$, the effects 
of the on-site and nearest-neighbor repulsion cancel each other. The dimensionless coupling 
constants  controlling the spin degrees of freedom (\ref{Param1}) are exactly 
the same as in the case of 
the PK model. Therefore we conclude that along the frustration line $U=2V $ an additional 
phase transition with increasing $W$ from the BO-SDW to the $\eta_{\pi}$-superconducting takes 
place with $W_{c}\simeq W_{c}(U=V=0) \simeq 2t$.  Numerical studies of this sector of the phase 
diagram are currently in progress and will be published elsewhere.

\section{Discussion and summary}

To summarize, we have presented the weak-coupling ground state phase diagram for 1D extended 
$U-V$ Hubbard  with pair-hopping in the case of a half-filled band. We have shown that the model 
has a very rich phase diagram which includes the singlet-superconducting phase, 
a metallic phase with 
dominating SS and TS instabilities and four different insulating phases corresponding to the Mott 
antiferromagnet, the CDW insulator, the bond-ordered CDW and the  bond-ordered SDW phase. 
In addition, we argued for the existence of a phase transition 
to the $\eta_{\pi}$-superconducting phase 
within the narrow stripe at $|U-2V| \ll 2t \leq W$. 
\vskip0.5cm
\begin{center}
ACKNOWLEDGEMENTS
\end{center}

GIJ gratefully acknowledges the kind hospitality at the Max Planck Institute for the
Physics of Complex Systems where part of this work has been done. He also acknowledges support 
by INTAS-Georgia grant N 97-1340. SS would like to thanks Marco Ameduri for a critical reading
of the manuscript.

\end{document}